\documentclass[11pt,twoside]{article}
\usepackage{asp2010}
\newcommand{\unit}[1]{\ensuremath{\:\mathrm{#1}}}
\newcommand{\muHz}{\unit{\mu Hz}}

\resetcounters

\begin{document}

\title{2D computations of g modes in fast rotating stars}
\author{J. Ballot,$^{1,2}$ F. Ligni\`eres,$^{1,2}$ V. Prat,$^{1,2}$ D. R. Reese$^3$ and M. Rieutord$^{1,2}$}
\affil{$^1$CNRS, IRAP, 14 avenue Edouard Belin, 31400 Toulouse, France}
\affil{$^2$Universit\'e de Toulouse, UPS-OMP, IRAP, 31400 Toulouse, France}
\affil{$^3$Observatoire de Paris, LESIA, CNRS, Universit\'e Pierre et Marie Curie, Universit\'e Denis Diderot, 5 place J. Janssen, 92195 Meudon, France }

\begin{abstract}
We present complete 2D computations of g modes in distorted polytropic models of stars performed with the Two-dimensional Oscillation Program (TOP). We computed low-degree modes ($\ell=1$ modes with radial order $n=-1\dots{-14}$, and $\ell=2,3$ modes with $n=-1\dots{-5}$ and $-16\dots{-20}$)
of a nonrotating model and followed them by slowly increasing the rotation rate up to 70~\% of the Keplerian break-up velocity. We use these computations to determine the domain of validity of perturbative methods up to the 3rd order. We study the evolution of the regularities of the spectrum and show quantitative agreement with the traditional approximation for not too large values of the ratio of the rotation rate to the pulsation frequency.
We also show the appearance of new types of modes, called ``rosette'' modes due to their spatial structure.
Thanks to the ray theory for gravito-inertial waves that we developed, we can associate these modes with stable periodic rays.
\end{abstract}

\section{Introduction}
Rapid rotators are typical among the main-sequence stars of intermediate or high masses \citep[\textit{e.g.},][]{Royer07}. Some stars exhibit oscillations due to waves trapped in the stellar interior. For some classes of pulsating stars, such as $\delta$~Sct or $\beta$~Cep, these oscillations are p (pressure) modes, whereas for $\gamma$~Dor or Slowly Pulsating B (SPB) stars, they are g (gravity) modes. For the present study, we focus on g modes, \textit{i.e.} low frequency modes driven by the buoyancy force.

To be able to model the oscillation spectrum of rotating stars, the effects of rotation must be carefully considered.
Rotation distorts stars under the effects of the centrifugal acceleration, and the Coriolis force can substantially modify the stellar oscillation properties. Complete computations are hard and delicate because it is a 2D  eigenvalue problem and it requires 2D models of stars.
To simplify the problem, the effects of rotation can be treated as a perturbation when the rotation rate, $\Omega$, is small relative to mode frequencies, $\omega$, and to the Keplerian break-up rotation rate, $\Omega_\mathrm{K}=(GM/R_\mathrm{eq}^3)^{1/2}$ ($M$ and $R_\mathrm{eq}$ are the mass and equatorial radius of the star, respectively).
Nevertheless, these hypotheses are not in the least verified for rapid rotators. 

We performed complete two-dimensional computations of g modes in distorted polytropic models of stars. After briefly presenting our computations (Sect.~\ref{sec:2dcomp}), we show the domains of validity of perturbative approaches (Sect.~\ref{sec:pert}), summarising results presented in \citet{Ballot10}. Then we discuss the evolution with rotation of the spectrum regularities (period spacing) and compare our results to those obtained with the traditional approximation (Sect.~\ref{sec:reg}). Finally, in Sect.~\ref{sec:eigen}, we discuss the spatial distribution of eigenmodes, especially the appearance of rosette modes.

\section{2D computations} \label{sec:2dcomp}
We consider fully radiative stars in this study. Since g modes are driven by the buoyancy force, they are evanescent in convective zones. SPB and $\gamma$~Dor stars have large radiative regions with a convective core, and a thin convective envelope for the latter. The effects of convective cores are not considered here.

 We approximate the equilibrium structure of rotating stars with self-gravitating uniformly-rotating polytropes, as already done for example in \citet{Lignieres06}. Polytropes are described in the co-rotating frame by the  following system of equations:
\begin{eqnarray}
p_o&=&K\rho_o^{1+1/\mu} \\
\nabla p_o &=& \rho_o \mathbf{g}_o \\
\Delta \psi_o &=& 4\pi G \rho_o
\end{eqnarray}
where $p_o$ is the pressure, $\rho_o$ the density, $\psi_o$ the gravitational potential, $K$ the polytropic constant, $\mu$ the polytropic index, $G$ the gravitational constant, and $\mathbf{g}_o$ the effective gravity, defined as
\begin{equation}
\mathbf{g}_o = - {\nabla}( \psi_o - \Omega^2 s^2 /2),
\end{equation} 
where $s$ the distance to the rotation axis.
Since the star is distorted by the centrifugal force, it is not spherical and we used a suited surface-fitting spheroidal system of coordinates based on \citet{Bonazzola98}. This equation system is numerically solved with the ESTER code \citep[Evolution STEllaire en Rotation, see][]{Rieutord05}. This is a spectral code using Chebychev polynomials in the pseudo-radial direction, and spherical harmonics in the horizontal one.
We computed models decomposed on spherical harmonics up to degree $\ell=32$, which ensures a sufficient accuracy for our purposes. 
The polytropic index is fixed to $\mu=3$, thereby approximating the structure of a fully radiative star.

In the co-rotating frame, the following equations govern the temporal evolution of small adiabatic inviscid perturbations of the equilibrium structure:
\begin{eqnarray}
\partial_t \rho &=& -\nabla\cdot (\rho_o \mathbf{v}) \label{eq:pert1}\\
\rho_o \partial_t \mathbf{v} &=&  -\nabla p + \rho \mathbf{g}_o -\rho_o \nabla \psi - 2\rho_o \mathbf{\Omega} \times \mathbf{v} \label{eq:pert2}\\
\partial_t p - c_o^2 \partial_t \rho &=& \frac{\rho_oN_o^2c_o^2}{g_o^2} \mathbf{v}\cdot \mathbf{g}_o \label{eq:pertenerg}\\
\Delta \psi &=& 4\pi G \rho \label{eq:pert4}
\end{eqnarray}
where $\rho$, $p$, $\mathbf{v}$, and $\psi$ are the Eulerian perturbations of density, pressure, velocity, and gravitational potential,
$c_o^2=\Gamma_1 p_o /\rho_o $ the adiabatic sound speed, and
$N_o^2 = \mathbf{g}_o \cdot (\nabla \ln \rho_o -\nabla \ln p_o / \Gamma_1 )$ the Brunt-V\"ais\"al\"a frequency. $\Gamma_1$ denotes the first adiabatic exponent.

We look for time-harmonic solutions of the system (\ref{eq:pert1})--(\ref{eq:pert4}), and we solve the obtained eigenvalue problem with the two-dimensional oscillation program \citep[TOP, see details in][]{Reese06}.
The equations are projected on the spherical harmonic basis $Y_{\ell}^m$. Due to the axisymmetry of the system, the projected equations are decoupled relatively to the azimuthal order $m$, but in contrast to the spherical nonrotating case, they are coupled for all degrees $\ell$ of the same parity. We compute modes with typically 20 spherical harmonics ($\ell_\mathrm{max}=40+|m|$).

We computed frequencies, $\omega_{n,\ell}^{(0)}$, of $\ell=1$ to 3 modes in a nonrotating polytrope (in this case, modes are described by one degree only). We considered $\ell=1$ modes with $n=-1$ to $-14$, and $\ell=2$ and 3 low-order ($n=-1$ to $-5$), and high-order ($n=-16$ to $-20$) modes.
We then followed the variation in frequency of each mode of degree $\ell$, azimuthal order $m$, and radial order $n$ by slowly increasing the rotation rate, step by step, from $0$ to $0.7\Omega_\mathrm{K}$. For a detailed description of the mode-following technique, we refer the readers to \citet{Ballot10}.

\section{Validity domain of perturbative methods} \label{sec:pert}
The effects of rotation on the oscillation modes can be treated as a perturbation where the rotation rate is 
the small parameter. A 1st-order correction has been proposed by \citet{Ledoux51}, 2nd-order by \citet{Saio81}, \citet{DziembowskiGoode92}, and \citet{Suarez06},  and 3rd-order terms have been developed by \citet{Soufi98}.
Within a perturbative approach, frequencies are developed as
\begin{equation}
\bar\omega_{n,\ell,m}^{pert}=\bar\omega_{n,\ell}^{(0)}+ mC_{n,\ell} \bar\Omega + 
(S^1_{n,\ell}+m^2S^2_{n,\ell}) \bar\Omega^2 +
m(T^1_{n,\ell}+m^2T^2_{n,\ell}) \bar\Omega^3 + {\cal O}(\bar\Omega^4) \label{eq:pertfin}
\end{equation}
The bar denotes the normalization $\bar\omega=\omega/\Omega_\mathrm{K}^{\mathrm{p}}$ and $\bar\Omega=\Omega/\Omega_\mathrm{K}^{\mathrm{p}}$ where $\Omega_\mathrm{K}^{\mathrm{p}}=(GM/R_\mathrm{p}^3)^{1/2}$ ($R_\mathrm{p}$ is the polar radius). We use this normalisation since the polar radius is expected to be a slowly varying function of $\Omega$ in real stars, as opposed to $R_\mathrm{eq}$.
We compute the perturbative coefficients $C_{n,\ell}$, $S^i_{n,\ell}$, and $T^i_{n,\ell}$ from our 2D computations as described in \citet{Reese06} and \citet{Ballot10}.

\begin{figure}[!htp]
\centering
\includegraphics[width=0.8\textwidth]{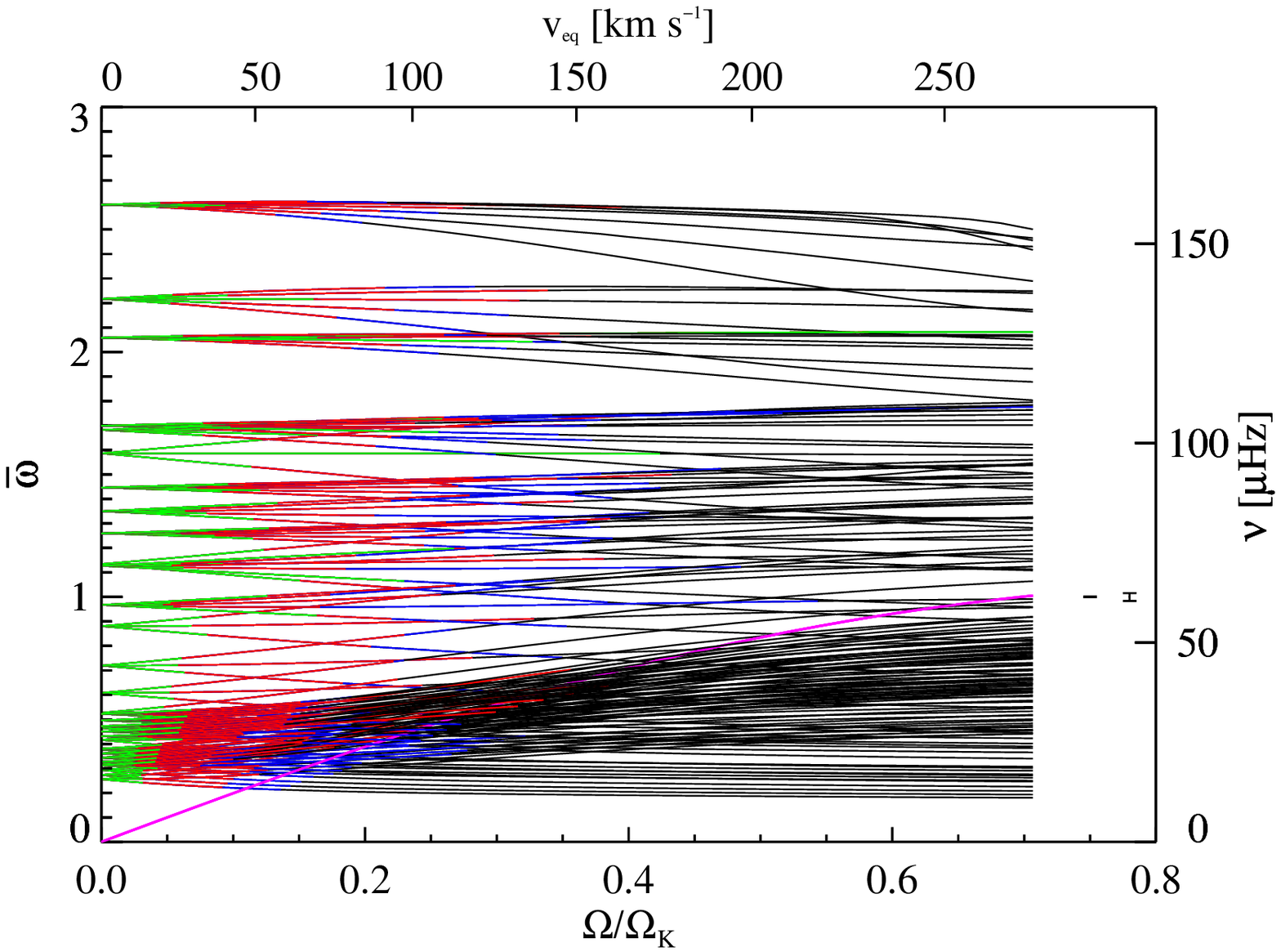}
\includegraphics[width=0.8\textwidth]{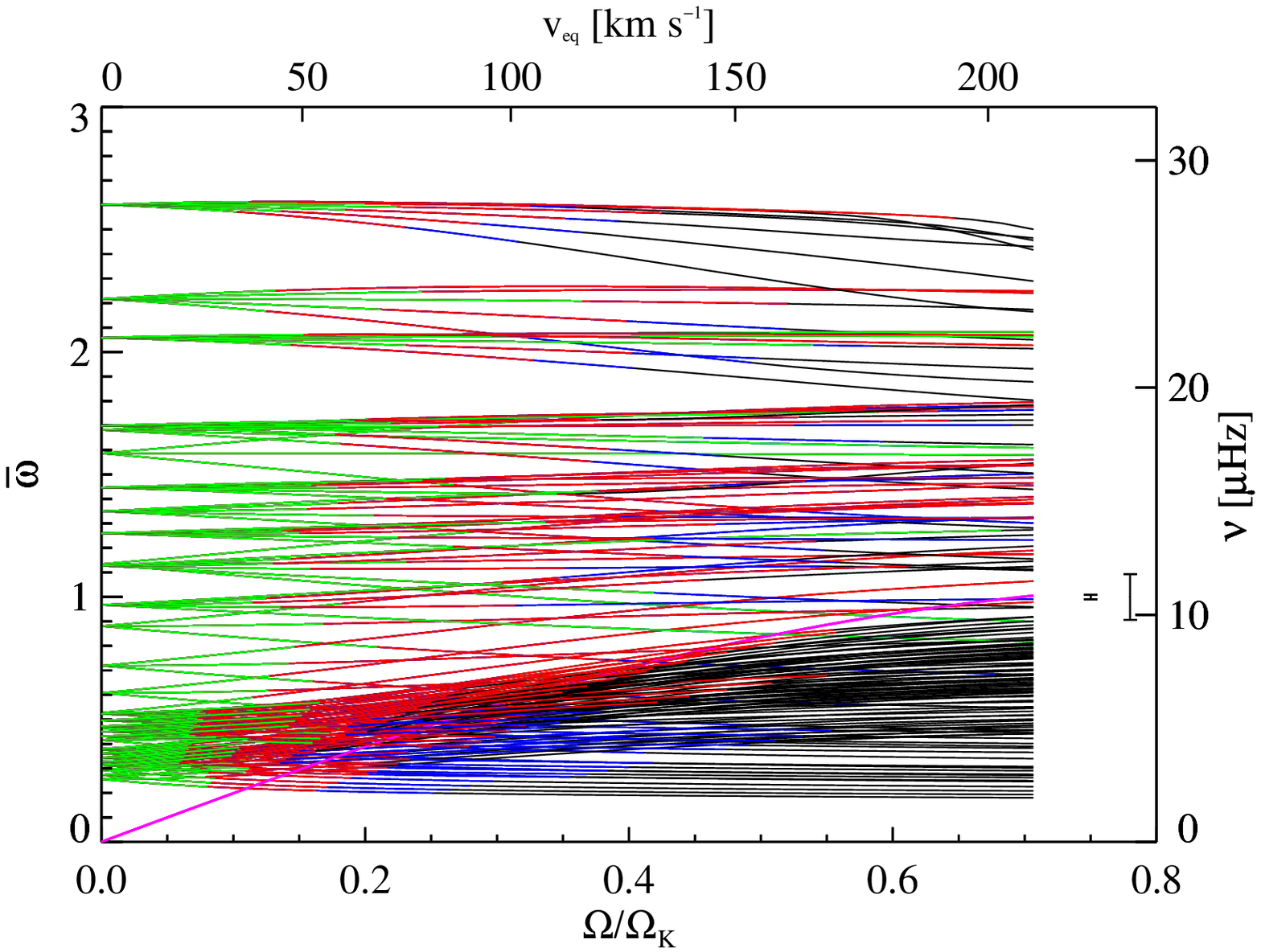}
\caption{Evolution of the frequencies of $\ell=1,2,3$ modes, computed in the co-rotating frame. 
Perturbative approximations have been tested for a typical $\gamma$~Dor \textit{(top panel)} and for a B star \textit{(bottom panel)}. The green/red/blue parts of curves indicate that 1st/2nd/3rd order is sufficient to reproduce complete calculations within an error $\delta\nu =0.1\muHz$. Error bars on the right-hand side of each panel show $\delta\nu$ and $10\times\delta\nu$. The magenta lines indicate $\omega=2\Omega$. For each plot, the bottom x-axis and left y-axis show dimensionless units, whereas the top x-axis and right y-axis show physical units.\label{fig:pert}}
\end{figure}

From these coefficients we calculate mode frequencies with the 1st to 3rd-order perturbative approximations for rotation rates ranging from $\Omega=0$ to $0.7\Omega_\mathrm{K}$ and compared them to complete computations.
To define the domains of validity of perturbative approaches, we fix the maximal departure, $\delta\bar\omega$, allowed between the perturbed frequencies, $\bar\omega^{pert}_{n,\ell,m}$, and the exact ones, $\bar\omega_{n,\ell,m}$. 
For each mode and each approximation order, we define the domain of validity $[0,\Omega_v]$, such that $\forall \Omega<\Omega_v\ |\bar\omega^{pert}_{n,\ell,m}(\Omega)- \bar\omega_{n,\ell,m}(\Omega)| < \delta\bar\omega$. 
The precision of the observed frequencies, $\delta \nu$, is related to the normalized error, $\delta\bar\omega$, through $\delta\bar\omega=\delta\nu/\nu_g$ with $\nu_g= \Omega_\mathrm{K}/(2\pi)$.
We thus display in Fig.~\ref{fig:pert} the domains of validity of the perturbative approximations for a frequency precision $\delta\nu=0.1\unit{\mu Hz}$ (spectral resolution after a hundred days) and for two types of stars with different dynamical frequencies:
a typical $\gamma$~Dor star ($\nu_g=61\muHz$), and a typical B star ($\nu_g=11\muHz$). Notice that these two plots can also be seen as the domains of validity of the perturbative approximations for a given star (\textit{e.g.}, a $\gamma$~Dor) star at two frequency precisions (\textit{e.g.}, 0.1 and 0.5\muHz).
The domains of validity obviously extend to higher rotation rates in the bottom panel.

We observe distinct behaviours in the high- and low-frequency ranges.
In the high-frequency range, 2nd-order perturbative methods give satisfactory results up to $\sim$100\unit{km\,s^{-1}} for $\gamma$~Dor stars
and up to $\sim$150\unit{km\,s^{-1}} for B stars. The 3rd-order terms improve the results and increase the domains of validity by a few tens of
\unit{km\,s^{-1}}. These results are to be contrasted with those found for p modes where
the domains of validity are restricted to lower rotation rates (for $\delta$~Sct stars, which are similar to $\gamma$~Dor, \citet{Reese06} find $\sim$50--70\unit{km\,s^{-1}} as a limit for perturbative methods) and where 3rd-order terms bring little improvement.
The rather good performance of perturbative methods at describing high-frequency g modes 
indicates in particular that the 2nd-order term gives a reasonable description of the centrifugal distortion.
This might be surprising considering the significant distortion of the stellar surface ($R_\mathrm{eq}/R_\mathrm{p} = 1.08$ at $\Omega = 0.4 \Omega_\mathrm{K}$). 
Actually, the energy of g modes is concentrated in the inner part of the star where 
the deviations from sphericity remain small.  As a result, g modes ``detect'' a much weaker distortion than p modes, thereby making them amenable to a perturbative description.
A particular feature that induces a strong deviation from the perturbative method concerns
mixed pressure-gravity modes that arise as a consequence of the centrifugal modification of the stellar structure.
For example, we found that, above a certain rotation rate, the $\ell=3,n=-1$ mode becomes a mixed mode with a p-mode character
in the outer low-latitude region associated with a drop in the Brunt-V\"ais\"al\"a frequency. 

However at low frequency, the domains of validity of perturbative methods are strongly reduced:
for $\gamma$~Dor stars, 2nd-order perturbative methods are only valid below $\sim$50\unit{km\,s^{-1}}.
Indeed, perturbative methods fail to recover the correct frequencies in the inertial regime $\omega<2\Omega$ (delimited by a magenta curve).
In particular, we observe that increasing the tolerance $\delta\bar\omega$ between
the top and bottom panels brings very little improvement in the inertial domain. As detailed in Sect.~\ref{ssec:inert}, this can be attributed to a change in the nature of modes in this regime.

\section{Regularities: Period spacing} \label{sec:reg}
\begin{figure}[!ht]
\centering
 \includegraphics[width=0.8\textwidth]{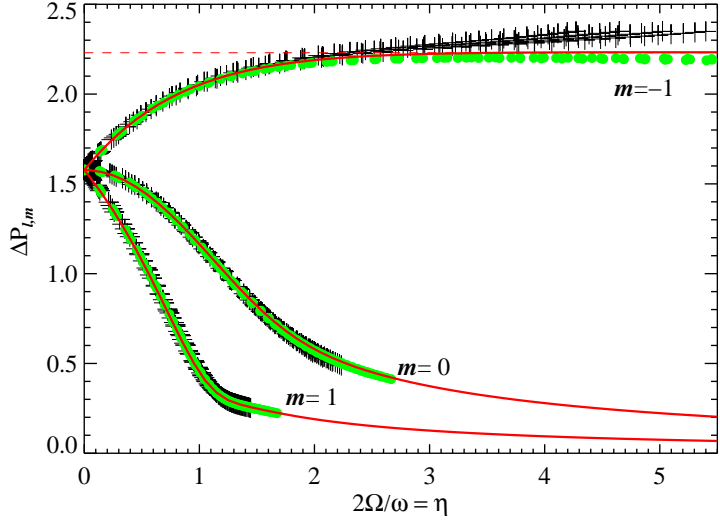}
\caption{Period spacings for $\ell=1$ modes as a function of $\eta$. Pluses correspond to complete computations, dots to computations on a spherical model with a full treatment of the Coriolis force. Modes with $n=-10$ to $-14$ are used. Solid lines correspond to the asymptotic relation obtained within the TA.\label{fig:dp}}
\end{figure}

For a nonrotating star, the period spacing $\Delta P_{n,\ell}=P_{n+1,\ell,0}-P_{n,\ell,0}$, where $P_{n,\ell,m}=2\pi/\omega_{n,\ell,m}$, is known to be asymptotically independent of $n$, in the absence of strong structure gradients. More precisely, $\Delta P_{n,\ell}\approx\Delta P_{\ell}=P_o/\sqrt{\ell(\ell+1)}$, where $P_o = 2\pi^2 \left(\int N_o/r\, \mathrm{d}r\right)^{-1}$ only depends on the structure of the star \citep{Tassoul80}.
For a rotating star this is not true. $\Delta P_{n,\ell,m}$ also depends on $\Omega$, $n$, and $m$. Nevertheless, for a given $m$, by considering sufficiently high values of $n$ ($n\gtrsim10$) we realise that the period spacing depends mainly on a reduced parameter, $\eta = 2\Omega/\omega$. Results for $\ell=1$ are displayed in Fig.~\ref{fig:dp}.

This parameter $\eta$ appears naturally within the so-called traditional approximation (TA). Commonly used in geophysics \citep{Eckart60}, it consists in assuming a spherical symmetry for the star and neglecting the tangential component of the rotation vector in Eq.~\ref{eq:pert2}. By doing this, the problem becomes separable in the radial and latitudinal coordinates. 
The equation governing the radial dependence
of the mode is similar to the nonrotating problem with $\ell(\ell+1)$ terms replaced by $\lambda_{\ell,m}(\eta)$, where $\lambda_{\ell,m}$ are the eigenvalues 
of the  Laplace's tidal eigenvalue problem in latitudinal direction \citep[\textit{e.g.},][]{Unno89}.
The TA has been used for computing g-mode frequencies in stars \citep[\textit{e.g.},][]{Berthomieu78,LeeSaio87,LeeSaio97,Townsend03}. 
Within the TA, for sufficiently high values of $n$, $P_{n,\ell,m}\approx nP_o/\sqrt{\lambda_{\ell,m}(\eta)}$. We  then derive the spacings
\begin{equation}
 \Delta P 
\approx \Delta P(\eta)
\approx \frac{P_o}{\sqrt{\lambda_{\ell,m}}\left(1+\frac{1}{2}\eta\frac{\lambda'_{\ell,m}}{\lambda_{\ell,m}}\right)}.\label{eq:dpta}
\end{equation}
The star is assumed to be spherical for the TA but spheroidal for the complete computations. Thus, in order to compare the results, we decided to consider that the polar radius of the distorted models is equal to the radius of the spherical model.
The relations (\ref{eq:dpta}) are plotted in Fig.~\ref{fig:dp} for $\ell=1$. They agree very well with our complete computations. We only see a departure between the asymptotic TA and exact frequencies for $m=-\ell$ modes for high values of $\eta$. For $\ell=2$ and 3, we get very similar results (not shown).
To investigate the origin of this discrepancy we performed complementary computations. We computed $\ell=1$ g-mode frequencies with TOP, without any simplification to the Coriolis term, but using a spherical model. The resulting period spacings are plotted with dots in Fig.~\ref{fig:dp}. They agree well with the TA even for $m=-\ell$, which demonstrates that the remaining difference originates in the centrifugal distortion of the 2D models.

\section{Spatial structure of the eigenmodes} \label{sec:eigen}
\subsection{Inertial domain $\omega<2\Omega$ ($\eta>1$)} \label{ssec:inert}
\begin{figure}[!ht]
\includegraphics[height=.4\textwidth]{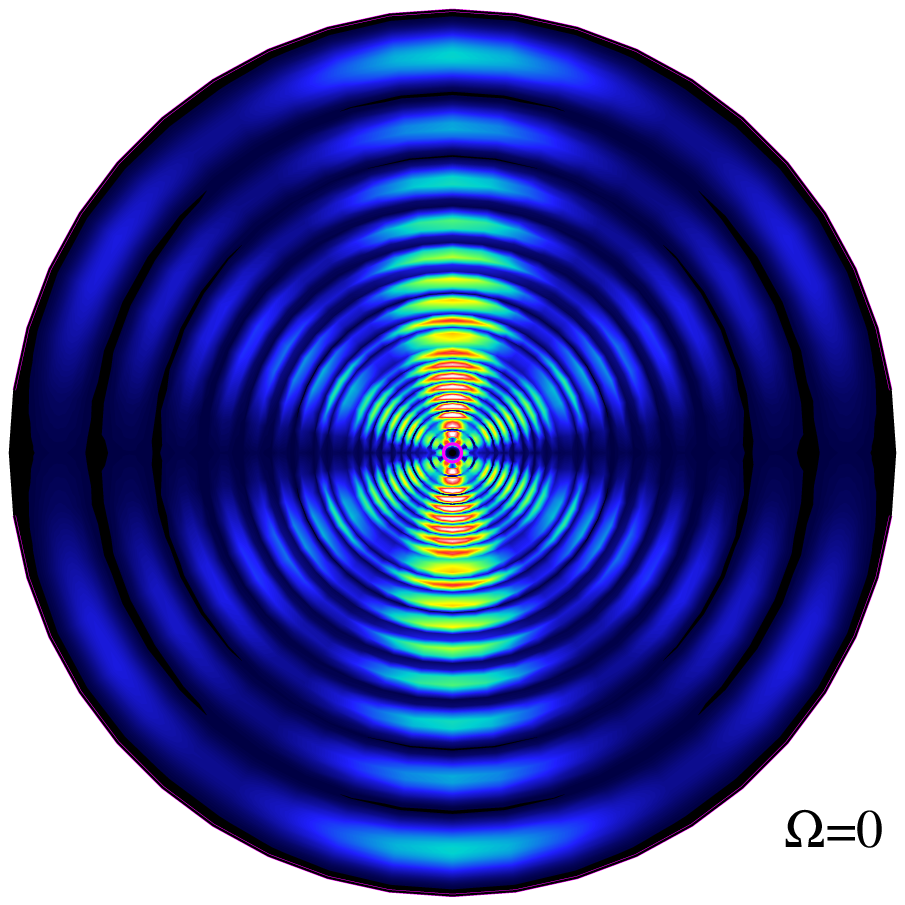}%
\includegraphics[height=.4\textwidth]{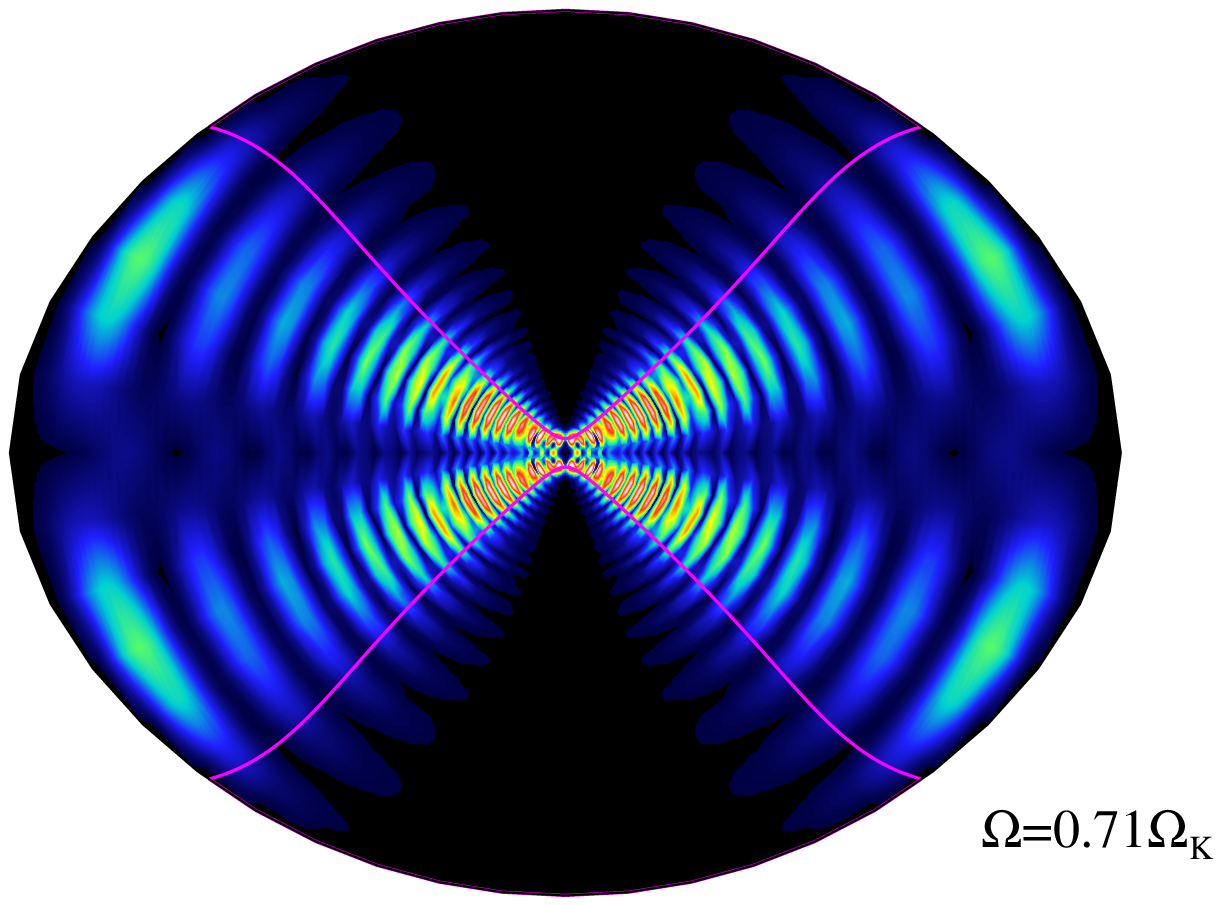}
\caption{\textit{(Left)} Meridional distribution of kinetic energy of the g mode $(\ell=3,m=-1,n=-18)$ in a nonrotating star. \textit{(Right)} The same for $\Omega=0.7\Omega_\mathrm{K}$. Magenta lines indicate the critical surface $\Gamma=0$. See also \citet{Ballot10}.\label{fig:forbid}}
\end{figure}

In Sect.~\ref{sec:pert}, we suggest that the failure of the perturbative method in the inertial regime is due to a change in the mode nature. In this regime, we observed changes in the mode cavity that are apparently not taken into account by the perturbative method. Modes in the inertial regime do not explore the polar region 
and the angular size of this forbidden region increases with $\eta$. This is illustrated in Fig.~\ref{fig:forbid} for a particular mode. Such a drastic change
in the shape of the resonant cavity has a direct impact on the associated mode frequency.
As perturbative methods ignore this effect, they cannot provide an accurate approximation of the frequencies in this regime.

This interpretation is supported by the analytical expression of the forbidden region determined by \citet{Dintrans00} for gravito-inertial modes. Indeed, for frequencies $\omega < 2\Omega$, the Coriolis force becomes a restoring force, and modes become mixed gravity-inertial modes. With a spherical model, and within the anelastic and Cowling approximations, they have shown that gravito-inertial waves only propagate in the region where
\begin{equation}
\Gamma = r^2 \omega^2 [N_o^2 +(2\Omega)^2 -\omega^2]
- (2\Omega N_o z)^2 > 0. \label{eq:crit}
\end{equation}
This implies that, when $\eta>1$, a critical latitude $\theta_c=\arcsin(1/\eta)$ appears above which waves cannot propagate.
Even though this expression does not strictly apply to our nonspherical geometry, we have superimposed 
the critical surfaces, $\Gamma=0$, over the energy distributions of our eigenmodes (Fig.~\ref{fig:forbid}).
Without rotation, there is only a small circle close to the centre, corresponding to the classical turning point $\omega=N_o$. For the mode with $\omega<2\Omega$, the polar forbidden region delineated by $\Gamma=0$ agrees pretty well with the energy distribution of the mode.

In real stars, which have convective cores, the gravito-inertial nature of modes could become even more crucial since the gravito-inertial waves propagate in convective regions when g modes cannot.

\subsection{Rosette modes}
\begin{figure}[!ht]
\includegraphics[height=.4\textwidth]{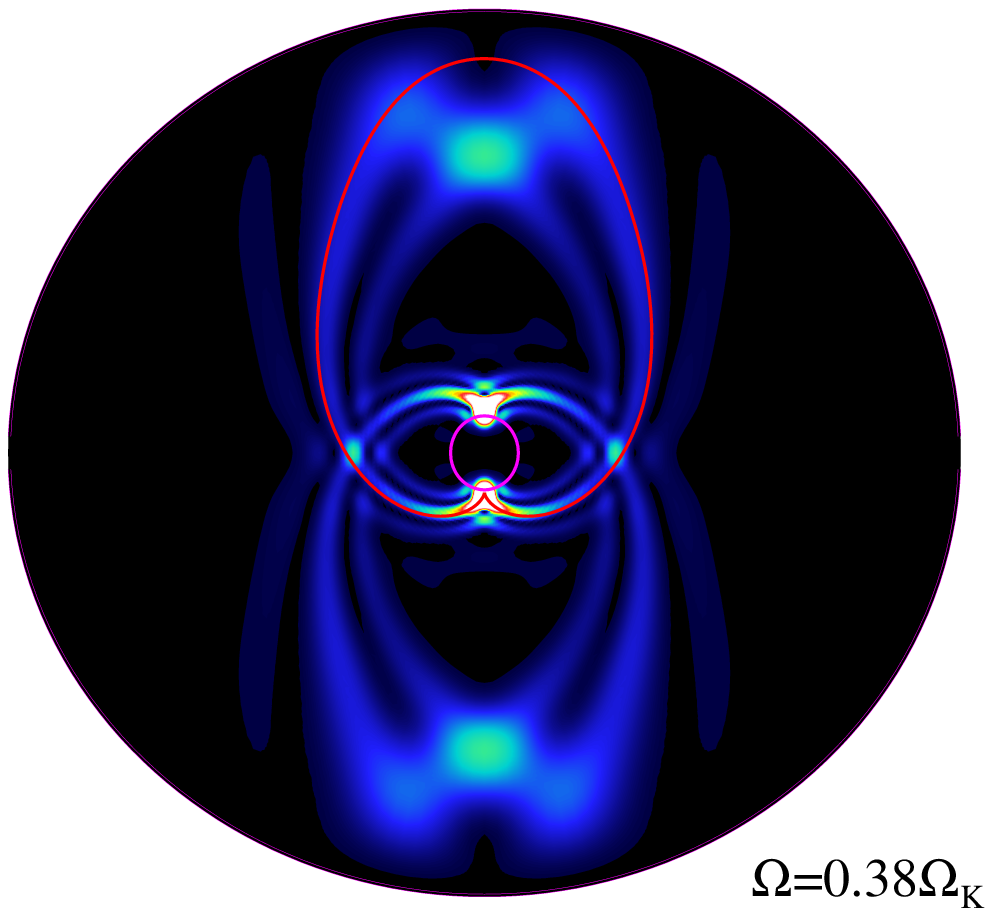}%
\includegraphics[height=.4\textwidth]{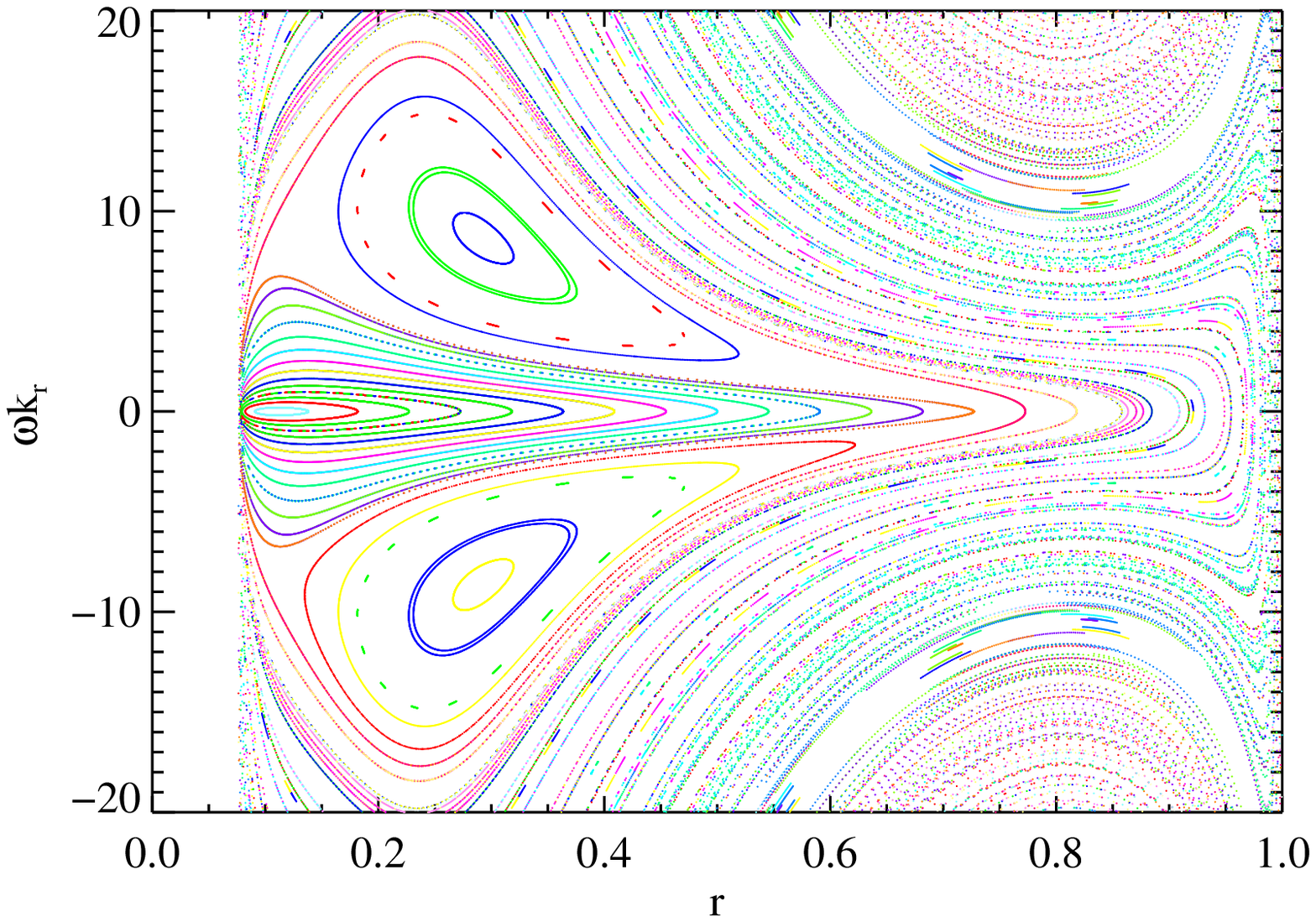}
\caption{\textit{(Left)} Meridional distribution of kinetic energy of the g mode $(\ell=3,m=0,n=-3)$ for $\Omega=0.38\Omega_\mathrm{K}$. The red line is a stable periodic orbit obtained from ray theory. \textit{(Right)} PSS corresponding to the same frequency ($\bar\omega=1.72$). The PSS is computed at $\theta=\pi/2$ (equatorial plane), $k_r$ denotes the radial component of the wave vector. \label{fig:rosette}}
\end{figure}

Even if they are not described by a unique spherical harmonics, modes with $\eta<1$ are generally strongly dominated by one degree. Hence, they look like ``distorted'' spherical harmonics (of course, avoided crossings sometimes create a strong coupling between modes leading to more complicated patterns).
However, for a few modes, such as $(\ell=3, m=0, n=-3)$, the mode structure changes quickly as the rotation rate increases (for this specific mode, changes are important at rotation rates as low as $0.1\Omega_\mathrm{K}$): the energy distribution focuses around a ``rosette'' pattern. Figure~\ref{fig:rosette} displays such a rosette mode.

Following the same approach as \citet{Lignieres08,Lignieres09} for p modes, 
Prat et al. (in preparation) have developed a ray theory for gravito-inertial modes in rotating stars. Using these developments, we computed the ray dynamics for the frequency of the $(\ell=3, m=0, n=-3)$ mode at $\Omega=0.38\Omega_\mathrm{K}$ ($\bar\omega=1.72$). The Poincar\'e surface of section (PSS) is plotted in Fig.~\ref{fig:rosette} \citep[for more details on the PSS and notions related to ray dynamics, see][]{Lignieres09}. We first see an unexplored region in the core ($r\lesssim 0.08$), corresponding to the region where $\omega<N_o$. The most striking features of this PSS are the two large islands formed around stable periodic orbits. One of these two stable orbits is superimposed over the mode distribution in Fig.~\ref{fig:rosette}, and 
 we see that the energy of the mode is distributed around it.
By exploring the spectrum with TOP in the vicinity of this mode (\textit{i.e.} $m=0$ modes with $\bar\omega\sim1.72$), we discover other rosette modes the energy of which is also distributed around the same rays, but with more nodes.

Thanks to ray theory, we also discover another family of rosette modes, around $\bar\omega=1.18$, corresponding to 2-period islands, that are also found with TOP.
All of these modes still exist even when the centrifugal distortion is neglected, \textit{i.e.} within a spherical approximation.

\acknowledgements The authors acknowledge the support from the French ANR through the SIROCO and ESTER projects. DRR gratefully acknowledges support from the CNES through a postdoctoral fellowship. This work was granted access to the HPC resources of CALMIP under the project P0107.

\bibliography{ballot}

\end{document}